\documentclass[preprint,aps ,nofootinbib]{revtex4}

\usepackage{parskip}
\usepackage{graphicx}
\usepackage{epsfig}
\usepackage{amsmath}
\usepackage{amsfonts}
\usepackage{amssymb}
\usepackage{bm}
\usepackage{color}
\usepackage{dcolumn}
\usepackage{hyperref}
\usepackage{multirow}
\usepackage{booktabs}
\usepackage{soul}
\usepackage{hyperref}
\usepackage{parskip}

\providecommand{\U}[1]{\protect\rule{.1in}{.1in}}

\newcommand{\f}{\begin{equation}}
\newcommand{\ff}{\end{equation}}
\newcommand{\fa}{\begin{eqnarray}}
\newcommand{\ffa}{\end{eqnarray}}
\newcommand{\ket}[1]{\vert#1\rangle}

\begin{document}
\title{Refined symmetry-resolved Page curve and charged black holes}
\author{Pan Li $^{1,2}$}
\email{lipan@ihep.ac.cn}
\author{Yi Ling $^{1,2}$}
\email{lingy@ihep.ac.cn} \affiliation{$^1$Institute of High Energy
Physics, Chinese Academy of Sciences, Beijing 100049, China\\ $^2$
School of Physics, University of Chinese Academy of Sciences,
Beijing 100049, China }

\begin{abstract}
The Page curve plotted by the typical random state approximation is not applicable to a system with conserved quantities, such as the evaporation process of a charged black hole during which the electric charge does not radiate out with a uniform rate  macroscopically. In this context the symmetry-resolved entanglement entropy may play a significant role in describing the entanglement structure of such a system. We attempt to impose constraints on microscopic quantum states to match with the macroscopic phenomenon of the charge radiation during black hole evaporation. Specifically, we consider a simple qubit system with conserved spin/charge serving as a toy model for the evaporation of charged black holes. We propose refined rules for selecting a random state with conserved quantities to simulate the distribution of charges during the different stages of evaporation, and obtain refined Page curves that exhibit distinct features in contrast to the original Page curve. We find the refined Page curve may have a different Page time and exhibit asymmetric behavior on both sides of the Page time. Such refined Page curves may provide more realistic description for the entanglement between the charged black hole and radiation during the process of evaporation. 
\end{abstract}

\maketitle
\flushbottom

\section{Introduction}
\label{sec:intro}
Quantum entanglement as one of the most prominent characteristics of a quantum system  has been shown to play an important role in many fields such as quantum information, quantum computation, condensed matter physics as well as black hole physics\cite{Horodecki:2009zz, Eisert:2008ur, Amico:2007ag, Laflorencie:2015eck,Hawking:1975vcx,Page:1993df,Page:1993wv}. It demonstrates the mysterious non-classical correlation between quantum subsystems. Entanglement entropy as an important measure of quantum entanglement has been extensively investigated. In particular, for a bipartite system Page \cite{Page:1993df} finds an important feature for the entanglement entropy between two subsystems, which now sometimes is referred to as ``Page's theorem". It states that the average entanglement entropy of the smaller subsystem over random pure states is very close to its maximal value, which is constrained by degrees of freedom in the subsystem. It means that typically the smaller subsystem is almost maximally entangled with the other subsystem. Inspired by this observation, Page originally noticed that this picture may be applicable to the famous black hole information loss paradox \cite{Hawking:1975vcx}\cite{Page:1993wv}. The core issue about this paradox is whether the evaporation process of a black hole due to Hawking radiation \footnote{For charged black holes, the Schwinger effect can also contribute to black hole evaporation \cite{Gibbons:1975kk}.}, which is a semi-classical result in quantum field theory over curved but classical spacetime, is fundamentally unitary at the complete quantum mechanical level. Currently a complete quantum theory of gravity has not been established yet and the microscopic description  of a black hole  with quantum states are unknown. Nevertheless, Page suggests that one might choose random states as the approximation of black hole states during the evaporation process\footnote{Recently some work \cite{Shenker:2013pqa, Cotler:2016fpe, Maldacena:2015waa} advocates that a black hole in some aspects behaves as a chaotic system, implying that the random approximation seems reasonable.}. Under the condition that the black hole evaporation process is unitary, it is believed that the Page's theorem is applicable to a system composed of the black hole and its radiation, and in principle one may find that the typical entanglement entropy of the radiation subsystem as the function of its size should follow the so-called Page curve. Such a picture proposed by Page stimulates a lot of work to further understand the black hole information paradox by investigating the entanglement between the black hole and radiation. In particular, since the island paradigm was proposed in the holographic approach\cite{Penington:2019npb,Almheiri:2019psf,Almheiri:2019hni}, reproducing the Page curve for the evaporation process of the black hole by holography has been becoming a central task in order to argue that the information would be released in the later half stage of radiation.

Nevertheless, we know the approximation with totally random states sometimes is not precise enough to describe the evolution of quantum entanglement in practice, since many physically relevant systems usually have conserved charges, such as energy, momentum, and electric charge. In a system with conserved charges, the reduced Hilbert space is composed of quantum states which are subject to the conservation of charges, thus do not equal to the tensor product of the Hilbert spaces \cite{Casini:2013rba}\cite{Ma:2015xes}, each of which is defined on individual subsystem separately. However, in this case one can still compute the entanglement entropy between two subsystems based on the reduced Hilbert space, which is the so-called ``symmetry-resolved (SR) entanglement entropy" \cite{Goldstein:2017bua}\cite{Xavier:2018kqb}, and has been widely investigated in many theoretical aspects \cite{bonsignori2019symmetry,belin2013holographic,caputa2016charged,dowker2016,dowker2017,cornfeld2018imbalance,
barghathi2018renyi,barghathi2019operationally,feldman2019dynamics,cornfeld2019entanglement,
fraenkel2020symmetry,calabrese2020full,monkman2020operational,azses2020symmetry,azses2020identification,
barghathi2020theory,tan2020particle,murciano2020symmetry,turkeshi2020entanglement,kiefer2020bounds,
murciano2020symmetry_2D,kiefer2020evidence,capizzi2020symmetry,murciano2020entanglement,horvath2020symmetry,
bonsignori2020boundary,estienne2021finite,murciano2021symmetry,
chen2021symmetry,zhao2021symmetry,weisenberger2021symmetry,
capizzi2021symmetry,hung2021entanglement,calabrese2021symmetry,
horvath2021u,azses2021observation,kiefer2021slow,shachar2021entanglement,parez2021exact,parez2021quasiparticle,ma2022symmetric,
oblak2022equipartition,zhao2022symmetry,ares2022symmetry,jones2022symmetry,
horvath2022branch,chen2022charged,ghasemi2022universal,scopa2022exact,parez2022dynamics,chen2022negativityboson,multicharged,goldstein2022,di2023boundary,Gaur:2022sjf,Gaur:2023yru} and even in experiments \cite{Lukin_2019}\cite{10.21468/SciPostPhys.12.3.106}. 
It is important to notice that when computing the average SR entanglement entropy, the original Page's theorem in general is not applicable. So it is pretty interesting to investigate the Page curve for SR entanglement entropy (SR Page curve) and to compare it with the page curve of systems without conserved charges. Previously some relevant work on this topic can be found in \cite{Murciano:2022lsw, Lau:2022hvc}.

It is well known that a stationary black hole is usually characterized by three conserved quantities classically, namely the mass $E$, the angular momentum $J$, and the electric charge $Q$.  When considering the evaporation of such a black hole, it is also quite natural to assume that these three quantities are conserved during the process of evaporation. Therefore, one may apply SR entanglement entropy to describe the entanglement between the black hole and the radiation. In principle, a refined Page curve may be obtained with a similar method introduced by Page, except that one just considers the average value of entanglement entropy over the non-factorized reduced Hilbert space  \cite{Bianchi:2019stn}. However, we point out that for the evaporation of a charged black hole, such naive calculation based on random states does not align with the semi-classical calculation of Hawking radiation and Schwinger effect \cite{Gibbons:1975kk}.  The key point is that the charge does not radiate out at the same rate as that of the mass \cite{Carter:1974yx,Gibbons:1975kk,Page:1976df,Page:1976ki,Hiscock:1990ex}.  Specifically, if we were to randomly select states in the entire Hilbert space with a fixed global charge number, the average charge number of the subsystem (radiation part) would increase linearly with the number of particles (at least in the case of the usually simple qubit model, see Figure. \ref{average charge}). However, the semi-classical calculation of Hawking radiation and Schwinger effect reveal that the evolution of electric charge $Q$, the mass $E$ and the angular momentum $J$ exhibits distinct behavior during the evaporation \cite{Carter:1974yx,Gibbons:1975kk,Page:1976df,Page:1976ki}. In general, for a black hole the rate of losing energy and angular momentum changes relatively slowly in the entire evaporation process \cite{Carter:1974yx}, but the radiation rate of electric charge $Q$ depends on the stage of the black hole which is specified by the relations of parameters \cite{Gibbons:1975kk}\cite{Hiscock:1990ex}. For different parameter values, the evaporation rate of the charge $Q$ varies dramatically, which will be described with details in Sec.\ref{sec:charged BH}. Consequently, the naive random model, even at a qualitative level, fails to capture the charge distribution during the course of evaporation for a charged black hole.  In this paper, aiming to simulate the evaporation of a charged black hole, we analyze the SR entanglement entropy in a qubit model with conserved charges and obtain various refined Page curves that reflect the different behavior of the black hole radiating its charge.  We will show that this model qualitatively captures the feature that the charge does not radiate out with a uniform rate during the evaporation of black holes.

This paper is organized as follows. In Sec.\ref{sec:page curve},  we review the calculation of the average entanglement entropy over totally random states in a system without conserved charges and the average SR entanglement entropy in a system with conserved charges. In Sec.\ref{sec:charged BH}, we introduce a qubit model for the evaporation of a charged black hole, and propose refined rules for selecting a random state with conserved quantity to simulate the distribution of charges during the different stages of evaporation, and then  obtain refined SR Page curves. In Sec.\ref{conclusion}, we present some discussions and outlooks on the work in future.

\section{The average entanglement entropy and Page curve based on random states }
\label{sec:page curve}
In this section, we will review the general consideration about the entanglement between two subsystems in a bipartite system which is described by a random pure quantum state. We firstly compute the average value of the entanglement entropy in a system without conserved charges, and then turn to the  SR entanglement entropy in a system with conserved charges. 
\subsection{The average entanglement entropy and Page curve in a system without conserved charges}

Given a bipartite system $A\cup B$ with the Hilbert space $\mathcal{H_{AB}}=\mathcal{H_A}\otimes \mathcal{H_B}$, where $\mathcal{H_A}$ and $\mathcal{H_B}$ are the Hilbert space of subsystem $A$ and $B$, respectively. Suppose the total system is described by a pure state $\ket{\psi}$, then the entanglement entropy of $A$ is defined by
\begin{equation}
    S_A=-\mathrm{Tr}(\rho_A\mathrm{ln}\rho_A),
\end{equation}
where $\rho_A$ is the reduced density matrix by tracing $B$. Suppose the dimension of $\mathcal{H_{A}}$ and $\mathcal{H_{B}}$ is $d(\mathcal{H_{A}})=d_{A}$, $d(\mathcal{H_{B}})=d_{B}$ respectively, then the dimension of  $\mathcal{H_{AB}}$ is $d(\mathcal{H_{AB}})=d_{AB}=d_{A}d_{B}$. Now we intend to compute the average value of the entanglement entropy  $\langle S_A\rangle$ for random states in $\mathcal{H_{AB}}$. We need first find the uniform measure in $\mathcal{H_{AB}}$. For this purpose we pick out an orthogonal basis $\{\ket{n}\}$ for a random state $\ket{\psi}= \sum_{n=1}^{d_{AB}} c_{n} \ket{n}$, and then the measure is just the uniform measure on the unit sphere of $\mathbb{C}^{d_{AB}}$, which is $d\mu(\psi)=\delta( \sum_{n=1}^{d_{AB}} |c_{n}|^2-1) \prod_{n=1}^{d_{AB}} dc_n d\Bar{c}_n$.

As a result, the average value of $S_A$ is obtained by integrating all the quantum states in Hilbert space $\mathcal{H_{AB}}$ 
\begin{equation}
    \langle S_A\rangle=\int -\mathrm{Tr}(\rho_A\mathrm{ln}\rho_A) d\mu(\psi).
\end{equation}
 One can also transform the integration variables into eigenvalues of $\rho_A$, where the details can be found in \cite{Bianchi:2019stn}.
The final result is 
\begin{align}
 \langle S_A \rangle&= \Psi(d_{AB}+1)-\Psi(d_{B}+1)-\frac{d_{A}-1}{2d_{B}} \\
&\simeq \ln d_A-\frac{d_{A}}{2 d_{B}} \text { for } 1 \ll d_A \leq d_B,
\end{align}
where $\Psi(x)=\Gamma'(x)/\Gamma(x)$ is the so called Digamma function. The above result indicates that for a bipartite system described by pure states, the smaller subsystem is almost maximally entangled with the other subsystem. So the Page curve, which plots the entanglement entropy as a function of the size of the subsystem,  will first increase with the size of the subsystem up to its maximal value at $d_A=d_B$, and then decrease with the size since for a pure system one always has $\langle S_A \rangle=\langle S_B \rangle $, which now is constrained by the size of the smaller subsystem $B$. Obviously, when the Hilbert space of the system is large enough, the Page time is located at $d_A=d_B$, and the curve exhibits a symmetric behavior on both sides of the Page time. We remark that this result is rooted at the uniform measure over the Hilbert space, thus does not depend on the details of evolution and in this sense may be treated as a model independent result.   

\subsection{The average SR entanglement entropy and SR Page curve in a system with conserved charges}

In a system with conserved charges, only those quantum states subject to these constraints are allowable, leading to a reduced Hilbert space which may be much smaller than the total Hilbert space.  For instance, a bipartite system contains a conserved charge $\hat{Q}$, then the total  Hilbert space can be decomposed into the direct sum of the eigenspace of $\hat{Q}$,
\begin{equation}
    \mathcal{H_{AB}=\sum_Q\mathcal{H_{AB}}}(Q).
\end{equation}

If the charge number $Q$ is fixed and conserved in a system, then we just need to care about one sector $\mathcal{H_{AB}}(Q)$. One immediate difference for  $\mathcal{H_{AB}}(Q)$ is that it can not be factorized into the tensor product of two Hilbert spaces of subsystems any more. Instead, it in general becomes  the direct sum of tensor product of the Hilbert spaces of subsystems with fixed charges
\begin{equation}
    \mathcal{H_{AB}}(Q)=\sum_{i=1}^{s} \mathcal{H_{A}}(q_i)\otimes\mathcal{H_B}(Q-q_i),
\end{equation}
where $s$ denotes the number of possible distributions of charges into two subsystems. In such a system, due to the fact that  $\mathcal{H_{AB}}(Q)\ne \mathcal{H_A}\otimes \mathcal{H_B}$, we need do more to figure out the uniform measure over the Hilbert space. By virtue of the direct sum structure of
$\mathcal{H_{AB}}(Q)$,  we may write a random state in $\mathcal{H_{AB}}(Q)$ as $\ket{\psi}= \sum_{i=1}^s\sqrt{p_i} \ket{\phi_i}$, with $p_i\ge0$ and $\sum_{i=1}^s p_i=1$, where $\ket{\phi_i} \in \mathcal{H_{A}}(q_i)\otimes\mathcal{H_B}(Q-q_i)$. As for each distribution $(q_i,Q-q_i)$, the corresponding Hilbert space has the form of tensor product. Therefore in this situation, the entanglement entropy of subsystem $A$  can be factorized into two parts
\begin{align}\label{eq:number}
S_A=&\sum_{i=1}^s p_i(q_i) S_{A}(q_i)-\sum_{i=1}^s p_i(q_i) \ln p_i(q_i).\
\end{align}
Here, $S_A(q_i)$ represents the entanglement entropy within the factorized Hilbert space $\mathcal{H_{A}}(q_i)\otimes\mathcal{H_B}(Q-q_i)$ for the state $\ket{\phi_i}$, which can be readily computed using the formula discussed in the previous subsection.

The uniform measure is also factorized into two part \cite{Bianchi:2019stn}
\begin{equation}
	\mathrm{d} \mu_{Q}(\psi)=\mathrm{d} \nu\left(p_{1}, \ldots, p_{s}\right) \prod_{i=1}^s \mathrm{d} \mu\left(\phi_{i}\right)\,,
\end{equation}
where $\mathrm{d} \nu\left(p_{1}, \ldots, p_{s}\right)$ is  the multivariate beta distribution \cite{Bianchi:2019stn}. After the average integration we obtain the final result for the average SR entanglement entropy which reads as \cite{Bianchi:2019stn} 
\begin{equation}
\textstyle \langle S_A \rangle_{\!{}_Q}=
\sum_{i=1}^s \!\frac{d_i}{d_{Q}} \big(\langle S_A(q_i) \rangle+\Psi(d_Q+1)  -\Psi(d_i +1)\big),
\end{equation}
where $d_i=d(\mathcal{H_{A}}(q_i)\otimes\mathcal{H_B}(Q-q_i))$, and $d_Q=\sum_{i=1}^sd_i$. Then in a parallel manner as described in the previous subsection we may obtain the SR Page curve as one changes the size of $A$ and $B$.

\section{A qubit model for charged black hole evaporation}
\label{sec:charged BH}
In this section we consider a simple qubit model with conserved charges to simulate the evaporation of a charged black hole, with the assumption that the process of evaporation is unitary. As we mentioned in Sec.\ref{sec:intro}, based on the analysis of Hawking radiation and Schwinger effect, charged black holes do not release their charge uniformly during evaporation \cite{Hiscock:1990ex} \footnote{Here ``uniform" refers to the condition that $\frac{dQ(t)}{dM(t)} $ is approximately constant, or at least it does not change dramatically.}. As a matter of fact, during evaporation the black hole may undergo different phases which depend on the black hole mass $M$ and electric charge $Q$, which has previously been revealed in \cite{Gibbons:1975kk}\cite{Hiscock:1990ex}\footnote{For black holes with mass $M<\frac{e^2}{m}\approx 10^{18}g\ll M_{\odot}$,  where $e$ and $m$ denote the charge and mass of a single electron respectively. Due to the presence of a strong electric field at the horizon, they can not possess even one electron for a reasonable length of time; for black holes with mass $M < \frac{e}{m^2} \approx 10^5M_{\odot}$, if the charge $Q$ is of a magnitude similar to $M$, the Schwinger effect will be strong, resulting in the rapid discharge of the black hole.}. Specifically, for the situations that are of primary interest in this paper, when $M> 2\times10^7M_{\odot}$, where $M_\odot$ denotes the mass of the Sun, the configuration space $(M,Q)$ may be divided into two regions according to the characteristics of evaporation: the ``mass dissipation zone" and the ``charge dissipation zone" \cite{Hiscock:1990ex}. When the charge-to-mass ratio of the black hole $Q/M$  is much less than one and $M$ is large, the black hole will be in the ``mass dissipation zone", then the black hole loses charges with a low rate $dQ(t)/dM(t)<Q(t)/M(t)$. Thus the ratio $Q(t)/M(t)$ becomes large as time $t$ goes by. On the other hand, when $Q/M$ is relatively close to one and $M$ is small, the black hole will be in the ``charge dissipation zone", then the black hole loses charges by Schwinger effect with a rate $dQ(t)/dM(t)>Q(t)/M(t)$ such that the ratio $Q(t)/M(t)$ becomes small with the evaporation. As a result, if a charged black hole starts to evaporate from a certain region within the  ``mass dissipation zone", its electric charge may remain nearly unchanged until more than half of its mass is lost. Only when the mass decreases to an order similar to the charge (in natural units), does significant charge release begin and the black hole enters the ``charge dissipation zone"\footnote{This process can be observed in Fig.2 of \cite{Hiscock:1990ex}}. Therefore, we conclude that the non-uniform release of charge is a common phenomenon during the evaporation of charged black holes.  Next, we apply a micro-level qubit model to simulate the evaporation of a charged black hole and propose refined rules to describe the charge release with a non-uniform rate, and investigate how the Page curve in this scenario differs from that evaluated in the case with completely random states.

We consider a qubit model which is composed of $N$ qubits to simulate a charged black hole radiating out particles. We divide the system into two subsystems $A$ and $B$, corresponding to the radiation and the charged black hole itself, respectively. For numerical analysis, we set the total number of qubits to be $N=20$, and require $N_A+N_B=20$ where $N_A$ and $N_B$ are the number of qubits in $A$ and $B$, respectively. Thus, the different partitions with $N_A$ from $0$ to $20$  represent the different stages of evaporation from the initial state to the final state. Roughly speaking, the number $N_B$ can be considered as a quantity analogous to the mass value $M$ of the black hole while $N_A$ is the energy of radiation. Alternatively, since the black hole loses its mass and the energy of radiation becomes larger during evaporation, the number of qubits $N_A$ may play a role of time as well. Next, we need to introduce a quantity to simulate the charge of a black hole. Just similar to the notion of spin, we assume each qubit may have a charge of either $+1$ or $-1$, so there is a 2-dimensional Hilbert space for each qubit. The total charge $Q$ of the system is defined as the eigenvalue of the operator $\hat{Q}=\sum_{i=1}^{N=20}{\sigma^i_z}$.  

As a typical pattern of evaporation, we consider the system to have a total charge of $Q=4$, which is aimed to simulate a black hole with an initially small ratio of the charge-to-mass. In the configuration with ($N=20$, $Q=4$), at the early stage of evaporation,  subsystem $B$ may fall into the “mass dissipation zone” due to $Q_B\ll N_B$. In this case, as a toy model, we stipulate that the subsystem $B$ does not release any charge until it shrinks to its half size, namely the half number of the total qubits. This condition implies that when $N_A$ is less than or equal to half of the total particles ($\frac{N}{2}$), the Hilbert space of radiation is $\mathcal{H}_A(q_i=0)$, and correspondingly $d(\mathcal{H_A}(q_i=0))=\binom{N_A}{\frac{N_A}{2}}$. On the other hand, the Hilbert space of the black hole is $\mathcal{H}_B(Q-q_i=4)$, and correspondingly $d(\mathcal{H_B}(Q-q_i=4))=\binom{N-N_A}{\frac{N-N_A+Q}{2}}$. After half of the black hole has evaporated, namely $N_A\ge \frac{N}{2}$, we consider the subsystem $B$ has entered the “charge dissipation zone” since $Q_B/N_B$ is large enough. For a qualitative description, we could require the charge to evaporate uniformly from the black hole afterward, therefore, the corresponding Hilbert space for radiation is $\mathcal{H_A}(q_i=\frac{Q}{(\frac{N}{2})}(N_A-\frac{N}{2}))$ with $N_A$ from $N/2$ to $N$. We remark that for a finite $N$, a technical problem may arise when $N_A$ jumps to $N$ with integer steps. Since for some $N_A$, the corresponding $q_i=\frac{Q}{(\frac{N}{2})}(N_A-\frac{N}{2})$ possibly is not an integer, then we have to skip this step or demand its integer part as $q_i$. When plotting the figure to illustrate the evolution of entanglement entropy with small $N$, it may cause visible imprecision. But we stress that such imprecision does not appear anymore in the thermodynamic limit, which requires $N_{th},Q_{th}\rightarrow \infty$, but keeping the radio $\frac{Q_{th}}{N_{th}}=\frac{Q}{N}=\frac{4}{20}$. We can just follow the calculations as above mentioned and then properly normalize the results.

\begin{figure}
\includegraphics[width=0.8\textwidth]{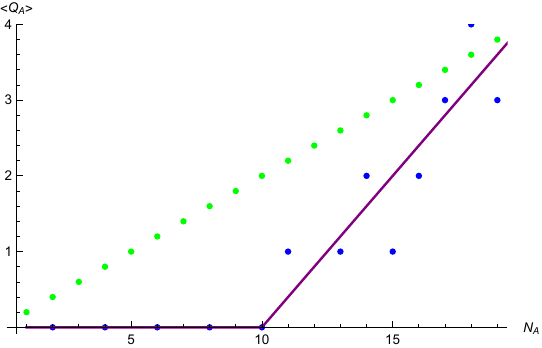}
\caption{The average charge number $\langle Q_A \rangle$ over random states with particle number $N_A$ in subsystems $A$ under different evaporation patterns. (i) The green dotted line represents the average value over random states in Hilbert space $\mathcal{H_{AB}}(Q=4)$ ; (ii) The blue dotted line represents the average value over random states in the refined Hilbert space  $\mathcal{H_{AB}}$; (iii) The purple solid line represents the average value over the random states in the refined Hilbert space $\mathcal{H_{AB}}$ in the thermodynamic limit, where the coordinates ($N_a$, $\langle Q_A \rangle$) correspond to ($N_a$, $\langle Q_A\rangle$)$\times\frac{20}{N_{th}} $ as a result of normalization. }
\label{average charge}
\end{figure}

\begin{figure}
\includegraphics[width=0.8\textwidth]{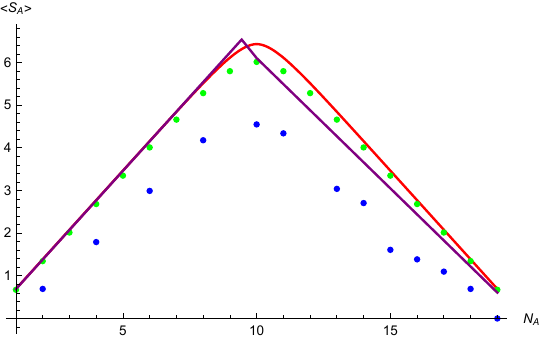}
\caption{The average entanglement entropy $\langle S_A \rangle$ over random states with particle number $N_A$ in subsystems $A$ under  different evaporation patterns. (i) The red solid line represents the average value over random states in the total Hilbert space $\mathcal{H_{AB}}=\mathcal{H_A}\otimes\mathcal{H_B}=\{\ket{-1},\ \ \ket{ 1}\}^{\otimes 20}$; (ii)\label{(i)} The green dotted line represents the average value over random states in the Hilbert space $\mathcal{H_{AB}}(Q=4)$ ; (iii) The blue dotted line represents the average value over random states in the refined Hilbert space $\mathcal{H_{AB}}$; (iv) The purple solid line represents the average value over random states in the refined Hilbert space $\mathcal{H_{AB}}$ in the thermodynamic limit, where the coordinates ($N_a$, $\langle S_A \rangle$) correspond to ($N_a$, $\langle S_A\rangle$)$\times\frac{20}{N_{th}} $ as a result of  normalization. }
\label{fig:page curve}
\end{figure}

Now we demonstrate our numerical results for $N=20$ and $Q=4$. For comparison we depict the average charge and average entanglement entropy as the function of $N_A$ for different patterns of evaporation in Figure (\ref{average charge}) and Figure (\ref{fig:page curve}). In Figure (\ref{average charge}), three different patterns for the charge  evaporation are demonstrated. The blue dotted line plots the numerical results for $N=20$, where we require no charge is released before $N_A\le \frac{N}{2}$, while after that moment at $N_A\ge \frac{N}{2}$, the charge is released with a constant rate. The purple solid line plots the expected charge profile in the thermodynamic limit, with the same rules. It is noticed that for $N=20$, the deviation from the case of thermodynamic limit with perfect constant rate for $N_A\ge \frac{N}{2}$  is visible, but the deviation would gradually disappear as $N$ get bigger. As a comparison, the green dotted line illustrates another pattern where the charge is released uniformly throughout the process of evaporation, which is obtained by computing the average value in  total Hilbert space $\mathcal{H_{AB}}(Q=4)$. Our main results are demonstrated in Figure (\ref{fig:page curve}), in which various Page curves are plotted for different patterns of evaporation. Evidently, in comparison with the Page curves with random state approximation, the refined Pages curves exhibit two prominent features. Firstly, the refined Page time may shift from the middle point $N_A=N/2$. Secondly, the refined Page curves exhibit asymmetric behavior on both sides of the Page time. Such features are understandable since in order to match the non-uniform charge release from the black hole, we apply different rules to  pick out different sectors of the total Hilbert space before and after the middle point $N_A=\frac{N}{2}$,  which disrupts the symmetry of the dimensions of Hilbert space before and after the middle point $N_A=\frac{N}{2}$. In addition, in the plot of the case with $N=20$ (the blue dotted line), we applied integer approximation to generate more data for drawing this plot. 

In the end of this section, we remark that we have only considered some typical patterns  of charged black hole evaporation with a specific set of parameters. Definitely one may perform a similar analysis for other patterns (which we have mentioned in the first paragraph of Sec III and footnote 4) with different discharge behavior.  One need impose different constraints on  original Hilbert space, and then pick up  reduced Hilbert space to match  different macroscopic patterns of evaporation. 

\section{Conclusion and Discussion}
\label{conclusion}
In this paper we have made the first step to  apply symmetry-resolved entanglement entropy to plot the Page curve for understanding the evaporation of charged black holes. Due to the non-uniform rate of discharge during evaporation, we should not simply evaluate  the average entanglement entropy over random states in a single Hilbert space for all time, 
as this would lead to a uniform charge release. As a toy model, we have computed the $SR$ entanglement entropy in a qubit system that simulates the charge distribution at different stages of evaporation for a charged black hole. After imposing restrictions on the Hilbert space of the system, we have obtained a reduced Hilbert space for each stage of the evaporation and plotted the refined $SR$ Page curve for this qubit system. We observe that the refined $SR$ Page curve exhibits two distinctive features compared to the random cases: it has a different Page time and displays asymmetric behavior on either side of the Page time.

Although the qubit system as a toy model is too simple to describe the quantum states of a genuine charged black hole, we stress that considering the entanglement structure of random states in the qubit model has grasped the spirit of the unitary evolution of a quantum chaotic system, which exhibits highly entangled behavior and may be viewed as an analogy of a black hole. The refined SR Page curve obtained in this paper matches with the macroscopic phenomenon of the discharge during black hole evaporation, and has shed light on understanding the release procedure of information from a charged black holes at a microscopic level. Definitely, many aspects of this model can be improved in future. Firstly, the analysis presented in this paper for charged black hole is somewhat at a qualitative level. For a more realistic black hole, various kinds of particles may evaporate, whose effects can not be ignored. The evaporation rate for different particles can be derived from semi-classical calculations of radiation  \cite{Gibbons:1975kk}.  Secondly, to simplify the analysis, we have only considered a linear relationship between charge and particle number release, while the actual quantitative relationship(e.g. $\frac{dQ_A(t)}{dt}$ and $\frac{dM_A(t)}{dt}$) would also depend on the specific calculations of evaporation details\cite{Page:1976df,Page:1976ki,Hiscock:1990ex}.  Thirdly, it is also interesting to consider a system with multiple conserved charges, such as energy and angular momentum. We expect the Page curve will differ quantitatively in such situations.

\section*{Acknowledgments}
We are  grateful to Kai Li, Wenbin Pan, Zhuoyu Xian, Zhangping Yu and Hongbao Zhang for their helpful discussions. We also appreciate the anonymous reviewer for pointing out some ambiguous expressions in our previous manuscript.
This work is supported in part by the Natural Science Foundation
of China under Grant No.~12035016 and 12275275. It is also supported by Beijing Natural Science Foundation under Grant No. 1222031, and  the Innovative Projects of Science and Technology No. E2545BU210 at IHEP.

\bibliographystyle{JHEP}
\bibliography{biblio.bib}

\providecommand{\href}[2]{#2}\begingroup\raggedright\begin{thebibliography}{10}

\bibitem{Horodecki:2009zz}
R.~Horodecki, P.~Horodecki, M.~Horodecki and K.~Horodecki, \emph{{Quantum
  entanglement}}, \href{https://doi.org/10.1103/RevModPhys.81.865}{\emph{Rev.
  Mod. Phys.} {\bfseries 81} (2009) 865}
  [\href{https://arxiv.org/abs/quant-ph/0702225}{{\ttfamily
  quant-ph/0702225}}].

\bibitem{Eisert:2008ur}
J.~Eisert, M.~Cramer and M.B.~Plenio, \emph{{Area laws for the entanglement
  entropy - a review}},
  \href{https://doi.org/10.1103/RevModPhys.82.277}{\emph{Rev. Mod. Phys.}
  {\bfseries 82} (2010) 277} [\href{https://arxiv.org/abs/0808.3773}{{\ttfamily
  0808.3773}}].

\bibitem{Amico:2007ag}
L.~Amico, R.~Fazio, A.~Osterloh and V.~Vedral, \emph{{Entanglement in many-body
  systems}}, \href{https://doi.org/10.1103/RevModPhys.80.517}{\emph{Rev. Mod.
  Phys.} {\bfseries 80} (2008) 517}
  [\href{https://arxiv.org/abs/quant-ph/0703044}{{\ttfamily
  quant-ph/0703044}}].

\bibitem{Laflorencie:2015eck}
N.~Laflorencie, \emph{{Quantum entanglement in condensed matter systems}},
  \href{https://doi.org/10.1016/j.physrep.2016.06.008}{\emph{Phys. Rept.}
  {\bfseries 646} (2016) 1} [\href{https://arxiv.org/abs/1512.03388}{{\ttfamily
  1512.03388}}].

\bibitem{Hawking:1975vcx}
S.W.~Hawking, \emph{{Particle Creation by Black Holes}},
  \href{https://doi.org/10.1007/BF02345020}{\emph{Commun. Math. Phys.}
  {\bfseries 43} (1975) 199}.

\bibitem{Page:1993df}
D.N.~Page, \emph{{Average entropy of a subsystem}},
  \href{https://doi.org/10.1103/PhysRevLett.71.1291}{\emph{Phys. Rev. Lett.}
  {\bfseries 71} (1993) 1291}
  [\href{https://arxiv.org/abs/gr-qc/9305007}{{\ttfamily gr-qc/9305007}}].

\bibitem{Page:1993wv}
D.N.~Page, \emph{{Information in black hole radiation}},
  \href{https://doi.org/10.1103/PhysRevLett.71.3743}{\emph{Phys. Rev. Lett.}
  {\bfseries 71} (1993) 3743}
  [\href{https://arxiv.org/abs/hep-th/9306083}{{\ttfamily hep-th/9306083}}].

\bibitem{Gibbons:1975kk}
G.W.~Gibbons, \emph{{Vacuum Polarization and the Spontaneous Loss of Charge by
  Black Holes}}, \href{https://doi.org/10.1007/BF01609829}{\emph{Commun. Math.
  Phys.} {\bfseries 44} (1975) 245}.

\bibitem{Shenker:2013pqa}
S.H.~Shenker and D.~Stanford, \emph{{Black holes and the butterfly effect}},
  \href{https://doi.org/10.1007/JHEP03(2014)067}{\emph{JHEP} {\bfseries 03}
  (2014) 067} [\href{https://arxiv.org/abs/1306.0622}{{\ttfamily 1306.0622}}].

\bibitem{Cotler:2016fpe}
J.S.~Cotler, G.~Gur-Ari, M.~Hanada, J.~Polchinski, P.~Saad, S.H.~Shenker
  et~al., \emph{{Black Holes and Random Matrices}},
  \href{https://doi.org/10.1007/JHEP05(2017)118}{\emph{JHEP} {\bfseries 05}
  (2017) 118} [\href{https://arxiv.org/abs/1611.04650}{{\ttfamily
  1611.04650}}].

\bibitem{Maldacena:2015waa}
J.~Maldacena, S.H.~Shenker and D.~Stanford, \emph{{A bound on chaos}},
  \href{https://doi.org/10.1007/JHEP08(2016)106}{\emph{JHEP} {\bfseries 08}
  (2016) 106} [\href{https://arxiv.org/abs/1503.01409}{{\ttfamily
  1503.01409}}].

\bibitem{Penington:2019npb}
G.~Penington, \emph{{Entanglement Wedge Reconstruction and the Information
  Paradox}}, \href{https://doi.org/10.1007/JHEP09(2020)002}{\emph{JHEP}
  {\bfseries 09} (2020) 002}
  [\href{https://arxiv.org/abs/1905.08255}{{\ttfamily 1905.08255}}].

\bibitem{Almheiri:2019psf}
A.~Almheiri, N.~Engelhardt, D.~Marolf and H.~Maxfield, \emph{{The entropy of
  bulk quantum fields and the entanglement wedge of an evaporating black
  hole}}, \href{https://doi.org/10.1007/JHEP12(2019)063}{\emph{JHEP} {\bfseries
  12} (2019) 063} [\href{https://arxiv.org/abs/1905.08762}{{\ttfamily
  1905.08762}}].

\bibitem{Almheiri:2019hni}
A.~Almheiri, R.~Mahajan, J.~Maldacena and Y.~Zhao, \emph{{The Page curve of
  Hawking radiation from semiclassical geometry}},
  \href{https://doi.org/10.1007/JHEP03(2020)149}{\emph{JHEP} {\bfseries 03}
  (2020) 149} [\href{https://arxiv.org/abs/1908.10996}{{\ttfamily
  1908.10996}}].

\bibitem{Casini:2013rba}
H.~Casini, M.~Huerta and J.A.~Rosabal, \emph{{Remarks on entanglement entropy
  for gauge fields}},
  \href{https://doi.org/10.1103/PhysRevD.89.085012}{\emph{Phys. Rev. D}
  {\bfseries 89} (2014) 085012}
  [\href{https://arxiv.org/abs/1312.1183}{{\ttfamily 1312.1183}}].

\bibitem{Ma:2015xes}
C.-T.~Ma, \emph{{Entanglement with Centers}},
  \href{https://doi.org/10.1007/JHEP01(2016)070}{\emph{JHEP} {\bfseries 01}
  (2016) 070} [\href{https://arxiv.org/abs/1511.02671}{{\ttfamily
  1511.02671}}].

\bibitem{Goldstein:2017bua}
M.~Goldstein and E.~Sela, \emph{{Symmetry-resolved entanglement in many-body
  systems}}, \href{https://doi.org/10.1103/PhysRevLett.120.200602}{\emph{Phys.
  Rev. Lett.} {\bfseries 120} (2018) 200602}
  [\href{https://arxiv.org/abs/1711.09418}{{\ttfamily 1711.09418}}].

\bibitem{Xavier:2018kqb}
J.C.~Xavier, F.C.~Alcaraz and G.~Sierra, \emph{{Equipartition of the
  entanglement entropy}},
  \href{https://doi.org/10.1103/PhysRevB.98.041106}{\emph{Phys. Rev. B}
  {\bfseries 98} (2018) 041106}
  [\href{https://arxiv.org/abs/1804.06357}{{\ttfamily 1804.06357}}].

\bibitem{bonsignori2019symmetry}
R.~Bonsignori, P.~Ruggiero and P.~Calabrese, \emph{Symmetry resolved
  entanglement in free fermionic systems},
  \href{https://doi.org/10.1088/1751-8121/ab4b77}{\emph{J. Phys. A: Math.
  Theor.} {\bfseries 52} (2019) 475302}.

\bibitem{belin2013holographic}
A.~Belin, L.-Y.~Hung, A.~Maloney, S.~Matsuura, R.C.~Myers and T.~Sierens,
  \emph{Holographic charged r{\'e}nyi entropies},
  \href{https://doi.org/10.1007/JHEP12(2013)059}{\emph{JHEP} {\bfseries 2013}
  (2013) 59}.

\bibitem{caputa2016charged}
P.~Caputa, M.~Nozaki and T.~Numasawa, \emph{Charged entanglement entropy of
  local operators},
  \href{https://doi.org/10.1103/PhysRevD.93.105032}{\emph{Phys. Rev. D}
  {\bfseries 93} (2016) 105032}.

\bibitem{dowker2016}
J.S.~Dowker{\emph{J. Phys. A: Math. Theor.} {\bfseries 49} (2016) 145401}.

\bibitem{dowker2017}
J.S.~Dowker{\emph{J. Phys. A: Math. Theor.} {\bfseries 50} (2017) 165401}.

\bibitem{cornfeld2018imbalance}
E.~Cornfeld, M.~Goldstein and E.~Sela, \emph{Imbalance entanglement: Symmetry
  decomposition of negativity},
  \href{https://doi.org/10.1103/PhysRevA.98.032302}{\emph{Phys. Rev. A}
  {\bfseries 98} (2018) 032302}.

\bibitem{barghathi2018renyi}
H.~Barghathi, C.M.~Herdman and A.~Del~Maestro, \emph{R\'enyi generalization of
  the accessible entanglement entropy},
  \href{https://doi.org/10.1103/PhysRevLett.121.150501}{\emph{Phys. Rev. Lett.}
  {\bfseries 121} (2018) 150501}.

\bibitem{barghathi2019operationally}
H.~Barghathi, E.~Casiano-Diaz and A.~Del~Maestro, \emph{Operationally
  accessible entanglement of one-dimensional spinless fermions},
  \href{https://doi.org/10.1103/PhysRevA.100.022324}{\emph{Phys. Rev. A}
  {\bfseries 100} (2019) 022324}.

\bibitem{feldman2019dynamics}
N.~Feldman and M.~Goldstein, \emph{Dynamics of charge-resolved entanglement
  after a local quench},
  \href{https://doi.org/10.1103/PhysRevB.100.235146}{\emph{Phys. Rev. B}
  {\bfseries 100} (2019) 235146}.

\bibitem{cornfeld2019entanglement}
E.~Cornfeld, L.A.~Landau, K.~Shtengel and E.~Sela, \emph{Entanglement
  spectroscopy of non-abelian anyons: Reading off quantum dimensions of
  individual anyons},
  \href{https://doi.org/10.1103/PhysRevB.99.115429}{\emph{Phys. Rev. B}
  {\bfseries 99} (2019) 115429}.

\bibitem{fraenkel2020symmetry}
S.~Fraenkel and M.~Goldstein, \emph{Symmetry resolved entanglement: exact
  results in 1d and beyond},
  \href{https://doi.org/10.1088/1742-5468/ab7753}{\emph{J. Stat. Mech.}
  {\bfseries 2020} (2020) 033106}.

\bibitem{calabrese2020full}
P.~Calabrese, M.~Collura, G.~Di~Giulio and S.~Murciano, \emph{Full counting
  statistics in the gapped xxz spin chain},
  \href{https://doi.org/10.1209/0295-5075/129/60007}{\emph{EPL} {\bfseries 129}
  (2020) 60007}.

\bibitem{monkman2020operational}
K.~Monkman and J.~Sirker, \emph{Operational entanglement of symmetry-protected
  topological edge states},
  \href{https://doi.org/10.1103/PhysRevResearch.2.043191}{\emph{Phys. Rev.
  Research} {\bfseries 2} (2020) 043191}.

\bibitem{azses2020symmetry}
D.~Azses and E.~Sela, \emph{Symmetry-resolved entanglement in
  symmetry-protected topological phases},
  \href{https://doi.org/10.1103/PhysRevB.102.235157}{\emph{Phys. Rev. B}
  {\bfseries 102} (2020) 235157}.

\bibitem{azses2020identification}
D.~Azses, R.~Haenel, Y.~Naveh, R.~Raussendorf, E.~Sela and E.G.~Dalla~Torre,
  \emph{Identification of symmetry-protected topological states on noisy
  quantum computers},
  \href{https://doi.org/10.1103/PhysRevLett.125.120502}{\emph{Phys. Rev. Lett.}
  {\bfseries 125} (2020) 120502}.

\bibitem{barghathi2020theory}
H.~Barghathi, J.~Yu and A.~Del~Maestro, \emph{Theory of noninteracting fermions
  and bosons in the canonical ensemble},
  \href{https://doi.org/10.1103/PhysRevResearch.2.043206}{\emph{Phys. Rev.
  Research} {\bfseries 2} (2020) 043206}.

\bibitem{tan2020particle}
M.T.~Tan and S.~Ryu, \emph{Particle number fluctuations, r\'enyi entropy, and
  symmetry-resolved entanglement entropy in a two-dimensional fermi gas from
  multidimensional bosonization},
  \href{https://doi.org/10.1103/PhysRevB.101.235169}{\emph{Phys. Rev. B}
  {\bfseries 101} (2020) 235169}.

\bibitem{murciano2020symmetry}
S.~Murciano, G.D.~Giulio and P.~Calabrese, \emph{Symmetry resolved entanglement
  in gapped integrable systems: a corner transfer matrix approach},
  \href{https://doi.org/10.21468/SciPostPhys.8.3.046}{\emph{SciPost Phys.}
  {\bfseries 8} (2020) 46}.

\bibitem{turkeshi2020entanglement}
X.~Turkeshi, P.~Ruggiero, V.~Alba and P.~Calabrese, \emph{Entanglement
  equipartition in critical random spin chains},
  \href{https://doi.org/10.1103/PhysRevB.102.014455}{\emph{Phys. Rev. B}
  {\bfseries 102} (2020) 014455}.

\bibitem{kiefer2020bounds}
M.~Kiefer-Emmanouilidis, R.~Unanyan, J.~Sirker and M.~Fleischhauer,
  \emph{Bounds on the entanglement entropy by the number entropy in
  non-interacting fermionic systems},
  \href{https://doi.org/10.21468/SciPostPhys.8.6.083}{\emph{SciPost Phys.}
  {\bfseries 8} (2020) 83}.

\bibitem{murciano2020symmetry_2D}
S.~Murciano, P.~Ruggiero and P.~Calabrese, \emph{Symmetry resolved entanglement
  in two-dimensional systems via dimensional reduction},
  \href{https://doi.org/10.1088/1742-5468/aba1e5}{\emph{J. Stat. Mech.}
  {\bfseries 2020} (2020) 083102}.

\bibitem{kiefer2020evidence}
M.~Kiefer-Emmanouilidis, R.~Unanyan, M.~Fleischhauer and J.~Sirker,
  \emph{Evidence for unbounded growth of the number entropy in many-body
  localized phases},
  \href{https://doi.org/10.1103/PhysRevLett.124.243601}{\emph{Phys. Rev. Lett.}
  {\bfseries 124} (2020) 243601}.

\bibitem{capizzi2020symmetry}
L.~Capizzi, P.~Ruggiero and P.~Calabrese, \emph{Symmetry resolved entanglement
  entropy of excited states in a cft},
  \href{https://doi.org/10.1088/1742-5468/ab96b6}{\emph{J. Stat. Mech.}
  {\bfseries 2020} (2020) 073101}.

\bibitem{murciano2020entanglement}
S.~Murciano, G.~Di~Giulio and P.~Calabrese, \emph{Entanglement and symmetry
  resolution in two dimensional free quantum field theories},
  \href{https://doi.org/10.1007/JHEP08(2020)073}{\emph{JHEP} {\bfseries 2020}
  (2020) 73}.

\bibitem{horvath2020symmetry}
D.X.~Horv{\'a}th and P.~Calabrese, \emph{Symmetry resolved entanglement in
  integrable field theories via form factor bootstrap},
  \href{https://doi.org/10.1007/JHEP11(2020)131}{\emph{JHEP} {\bfseries 2020}
  (2020) 131}.

\bibitem{bonsignori2020boundary}
R.~Bonsignori and P.~Calabrese, \emph{Boundary effects on symmetry resolved
  entanglement}, \href{https://doi.org/10.1088/1751-8121/abcc3a}{\emph{J. Phys.
  A: Math. Theor.} {\bfseries 54} (2020) 015005}.

\bibitem{estienne2021finite}
B.~Estienne, Y.~Ikhlef and A.~Morin-Duchesne, \emph{Finite-size corrections in
  critical symmetry-resolved entanglement},
  \href{https://doi.org/10.21468/SciPostPhys.10.3.054}{\emph{SciPost Phys.}
  {\bfseries 10} (2021) 54}.

\bibitem{murciano2021symmetry}
S.~Murciano, R.~Bonsignori and P.~Calabrese, \emph{Symmetry decomposition of
  negativity of massless free fermions},
  \href{https://doi.org/10.21468/SciPostPhys.10.5.111}{\emph{SciPost Phys.}
  {\bfseries 10} (2021) 111}.

\bibitem{chen2021symmetry}
H.-H.~Chen, \emph{Symmetry decomposition of relative entropies in conformal
  field theory}, \href{https://doi.org/10.1007/JHEP07(2021)084}{\emph{JHEP}
  {\bfseries 2021} (2021) 84}.

\bibitem{zhao2021symmetry}
S.~Zhao, C.~Northe and R.~Meyer, \emph{Symmetry-resolved entanglement in
  ads3/cft2 coupled to u (1) chern-simons theory},
  \href{https://doi.org/10.1007/JHEP07(2021)030}{\emph{JHEP} {\bfseries 2021}
  (2021) 30}.

\bibitem{weisenberger2021symmetry}
K.~Weisenberger, S.~Zhao, C.~Northe and R.~Meyer, \emph{Symmetry-resolved
  entanglement for excited states and two entangling intervals in ads3/cft2},
  \href{https://doi.org/10.1007/JHEP12(2021)104}{\emph{JHEP} {\bfseries 2021}
  (2021) 104}.

\bibitem{capizzi2021symmetry}
L.~Capizzi and P.~Calabrese, \emph{Symmetry resolved relative entropies and
  distances in conformal field theory},
  \href{https://doi.org/10.1007/JHEP10(2021)195}{\emph{JHEP} {\bfseries 2021}
  (2021) 195}.

\bibitem{hung2021entanglement}
L.Y.~Hung and G.~Wong, \emph{Entanglement branes and factorization in conformal
  field theory}, \href{https://doi.org/10.1103/PhysRevD.104.026012}{\emph{Phys.
  Rev. D} {\bfseries 104} (2021) 026012}.

\bibitem{calabrese2021symmetry}
P.~Calabrese, J.~Dubail and S.~Murciano, \emph{Symmetry-resolved entanglement
  entropy in wess-zumino-witten models},
  \href{https://doi.org/10.1007/JHEP10(2021)067}{\emph{JHEP} {\bfseries 2021}
  (2021) 67}.

\bibitem{horvath2021u}
D.X.~Horv{\'a}th, L.~Capizzi and P.~Calabrese, \emph{U (1) symmetry resolved
  entanglement in free 1+ 1 dimensional field theories via form factor
  bootstrap}, \href{https://doi.org/10.1007/JHEP05(2021)197}{\emph{JHEP}
  {\bfseries 2021} (2021) 197}.

\bibitem{azses2021observation}
D.~Azses, E.G.~Dalla~Torre and E.~Sela, \emph{Observing floquet topological
  order by symmetry resolution},
  \href{https://doi.org/10.1103/PhysRevB.104.L220301}{\emph{Phys. Rev. B}
  {\bfseries 104} (2021) L220301}.

\bibitem{kiefer2021slow}
M.~Kiefer-Emmanouilidis, R.~Unanyan, M.~Fleischhauer and J.~Sirker, \emph{Slow
  delocalization of particles in many-body localized phases},
  \href{https://doi.org/10.1103/PhysRevB.103.024203}{\emph{Phys. Rev. B}
  {\bfseries 103} (2021) 024203}.

\bibitem{shachar2021entanglement}
S.~Fraenkel and M.~Goldstein, \emph{Entanglement measures in a nonequilibrium
  steady state: Exact results in one dimension},
  \href{https://doi.org/10.21468/SciPostPhys.11.4.085}{\emph{SciPost Phys.}
  {\bfseries 11} (2021) 85}.

\bibitem{parez2021exact}
G.~Parez, R.~Bonsignori and P.~Calabrese, \emph{Exact quench dynamics of
  symmetry resolved entanglement in a free fermion chain},
  \href{https://doi.org/10.1088/1742-5468/ac21d7}{\emph{J. Stat. Mech.}
  {\bfseries 2021} (2021) 093102}.

\bibitem{parez2021quasiparticle}
G.~Parez, R.~Bonsignori and P.~Calabrese, \emph{Quasiparticle dynamics of
  symmetry-resolved entanglement after a quench: Examples of conformal field
  theories and free fermions},
  \href{https://doi.org/10.1103/PhysRevB.103.L041104}{\emph{Phys. Rev. B}
  {\bfseries 103} (2021) L041104}.

\bibitem{ma2022symmetric}
Z.~Ma, C.~Han, Y.~Meir and E.~Sela, \emph{Symmetric inseparability and number
  entanglement in charge-conserving mixed states},
  \href{https://doi.org/10.1103/PhysRevA.105.042416}{\emph{Phys. Rev. A}
  {\bfseries 105} (2022) 042416}.

\bibitem{oblak2022equipartition}
B.~Oblak, N.~Regnault and B.~Estienne, \emph{Equipartition of entanglement in
  quantum hall states},
  \href{https://doi.org/10.1103/PhysRevB.105.115131}{\emph{Phys. Rev. B}
  {\bfseries 105} (2022) 115131}.

\bibitem{zhao2022symmetry}
S.~Zhao, C.~Northe, K.~Weisenberger and R.~Meyer, \emph{Charged moments in w3
  higher spin holography},
  \href{https://doi.org/10.1007/JHEP05(2022)166}{\emph{JHEP} {\bfseries 2022}
  (2022) 1}.

\bibitem{ares2022symmetry}
F.~Ares, S.~Murciano and P.~Calabrese, \emph{Symmetry-resolved entanglement in
  a long-range free-fermion chain}, {\emph{arXiv:2202.05874} (2022) }.

\bibitem{jones2022symmetry}
N.G.~Jones, \emph{Symmetry-resolved entanglement entropy in critical
  free-fermion chains}, {\emph{arXiv:2202.11728} (2022) }.

\bibitem{horvath2022branch}
D.X.~Horvath, P.~Calabrese and O.A.~Castro-Alvaredo, \emph{Branch point twist
  field form factors in the sine-gordon model ii: Composite twist fields and
  symmetry resolved entanglement},
  \href{https://doi.org/10.21468/SciPostPhys.12.3.088}{\emph{SciPost Phys.}
  {\bfseries 12} (2022) 88}.

\bibitem{chen2022charged}
H.-H.~Chen, \emph{Charged renyi negativity of massless free bosons},
  \href{https://doi.org/10.1007/JHEP02(2022)117}{\emph{JHEP} {\bfseries 2022}
  (2022) 117}.

\bibitem{ghasemi2022universal}
M.~Ghasemi, \emph{Universal thermal corrections to symmetry-resolved
  entanglement entropy and full counting statistics}, {\emph{arXiv:2203.06708}
  (2022) }.

\bibitem{scopa2022exact}
S.~Scopa and D.X.~Horváth, \emph{Exact hydrodynamic description of
  symmetry-resolved rényi entropies after a quantum quench},
  {\emph{arXiv:2205.02924} (2022) }.

\bibitem{parez2022dynamics}
G.~Parez, R.~Bonsignori and P.~Calabrese, \emph{Dynamics of
  charge-imbalance-resolved entanglement negativity after a quench in a
  free-fermion model}, {\emph{J. Stat. Mech.} {\bfseries 2022} (2022) 053103}.

\bibitem{chen2022negativityboson}
H.-H.~Chen, \emph{Dynamics of charge imbalance resolved negativity after a
  global quench in free scalar field theory}, {\emph{arXiv:2205.09532} (2022)
  }.

\bibitem{multicharged}
F.~Ares, P.~Calabrese, G.~Di~Giulio and S.~Murciano{\emph{arXiv:2206.01534}
  (2022) }.

\bibitem{goldstein2022}
S.~Fraenkel and M.~Goldstein{\emph{arXiv:2205.12991} (2022) }.

\bibitem{di2023boundary}
G.~Di~Giulio, R.~Meyer, C.~Northe, H.~Scheppach and S.~Zhao, \emph{On the
  boundary conformal field theory approach to symmetry-resolved entanglement},
  {\emph{SciPost Physics Core} {\bfseries 6} (2023) 049}.

\bibitem{Gaur:2022sjf}
H.~Gaur and U.A.~Yajnik, \emph{{Charge imbalance resolved R\'enyi negativity
  for free compact boson: Two disjoint interval case}},
  \href{https://doi.org/10.1007/JHEP02(2023)118}{\emph{JHEP} {\bfseries 02}
  (2023) 118} [\href{https://arxiv.org/abs/2210.06743}{{\ttfamily
  2210.06743}}].

\bibitem{Gaur:2023yru}
H.~Gaur and U.A.~Yajnik, \emph{{Multi-charged moments and symmetry-resolved
  R\'enyi entropy of free compact boson for multiple disjoint intervals}},
  \href{https://doi.org/10.1007/JHEP01(2024)042}{\emph{JHEP} {\bfseries 01}
  (2024) 042} [\href{https://arxiv.org/abs/2310.14186}{{\ttfamily
  2310.14186}}].

\bibitem{Lukin_2019}
A.~Lukin, M.~Rispoli, R.~Schittko, M.E.~Tai, A.M.~Kaufman, S.~Choi et~al.,
  \emph{Probing entanglement in a many-body{\textendash}localized system},
  \href{https://doi.org/10.1126/science.aau0818}{\emph{Science} {\bfseries 364}
  (2019) 256}.

\bibitem{10.21468/SciPostPhys.12.3.106}
V.~Vitale, A.~Elben, R.~Kueng, A.~Neven, J.~Carrasco, B.~Kraus et~al.,
  \emph{{Symmetry-resolved dynamical purification in synthetic quantum
  matter}}, \href{https://doi.org/10.21468/SciPostPhys.12.3.106}{\emph{SciPost
  Phys.} {\bfseries 12} (2022) 106}.

\bibitem{Murciano:2022lsw}
S.~Murciano, P.~Calabrese and L.~Piroli, \emph{{Symmetry-resolved Page
  curves}}, \href{https://doi.org/10.1103/PhysRevD.106.046015}{\emph{Phys. Rev.
  D} {\bfseries 106} (2022) 046015}
  [\href{https://arxiv.org/abs/2206.05083}{{\ttfamily 2206.05083}}].

\bibitem{Lau:2022hvc}
P.H.C.~Lau, T.~Noumi, Y.~Takii and K.~Tamaoka, \emph{{Page curve and
  symmetries}}, \href{https://doi.org/10.1007/JHEP10(2022)015}{\emph{JHEP}
  {\bfseries 10} (2022) 015}
  [\href{https://arxiv.org/abs/2206.09633}{{\ttfamily 2206.09633}}].

\bibitem{Bianchi:2019stn}
E.~Bianchi and P.~Dona, \emph{{Typical entanglement entropy in the presence of
  a center: Page curve and its variance}},
  \href{https://doi.org/10.1103/PhysRevD.100.105010}{\emph{Phys. Rev. D}
  {\bfseries 100} (2019) 105010}
  [\href{https://arxiv.org/abs/1904.08370}{{\ttfamily 1904.08370}}].

\bibitem{Carter:1974yx}
B.~Carter, \emph{{Charge and particle conservation in black hole decay}},
  \href{https://doi.org/10.1103/PhysRevLett.33.558}{\emph{Phys. Rev. Lett.}
  {\bfseries 33} (1974) 558}.

\bibitem{Page:1976df}
D.N.~Page, \emph{{Particle Emission Rates from a Black Hole: Massless Particles
  from an Uncharged, Nonrotating Hole}},
  \href{https://doi.org/10.1103/PhysRevD.13.198}{\emph{Phys. Rev. D} {\bfseries
  13} (1976) 198}.

\bibitem{Page:1976ki}
D.N.~Page, \emph{{Particle Emission Rates from a Black Hole. 2. Massless
  Particles from a Rotating Hole}},
  \href{https://doi.org/10.1103/PhysRevD.14.3260}{\emph{Phys. Rev. D}
  {\bfseries 14} (1976) 3260}.

\bibitem{Hiscock:1990ex}
W.A.~Hiscock and L.D.~Weems, \emph{{Evolution of Charged Evaporating Black
  Holes}}, \href{https://doi.org/10.1103/PhysRevD.41.1142}{\emph{Phys. Rev. D}
  {\bfseries 41} (1990) 1142}.

\end{thebibliography}\endgroup
\end{document}